\documentclass[aps,twocolumn,PRB,reprint,superscriptaddress]{revtex4-1}
\pdfoutput=1
\usepackage{amsmath}
\usepackage{epsfig}
\usepackage{graphicx}
\usepackage{graphicx, color, epstopdf}
\usepackage{bm}
\usepackage{amssymb}
\usepackage{hyperref}
\usepackage{xcolor}
\usepackage{subfigure}
\begin{document}
\bibliographystyle{apsrev4-1}
\title{Aharonov-Bohm effect in three-dimensional higher-order topological insulators}
\author{Kun Luo}
\affiliation{National Laboratory of Solid State Microstructures and school of Physics, Nanjing University, Nanjing, 210093, China}

\author{Hao Geng}
\affiliation{National Laboratory of Solid State Microstructures and school of Physics, Nanjing University, Nanjing, 210093, China}
\affiliation{Collaborative Innovation Center of Advanced Microstructures, Nanjing University, Nanjing 210093, China}

\author{Li Sheng}
\affiliation{National Laboratory of Solid State Microstructures and school of Physics, Nanjing University, Nanjing, 210093, China}
\affiliation{Collaborative Innovation Center of Advanced Microstructures, Nanjing University, Nanjing 210093, China}

\author{Wei Chen}
\email{Corresponding author: pchenweis@gmail.com}
\affiliation{National Laboratory of Solid State Microstructures and school of Physics, Nanjing University, Nanjing, 210093, China}
\affiliation{Collaborative Innovation Center of Advanced Microstructures, Nanjing University, Nanjing 210093, China}

\author{D. Y. Xing}
\affiliation{National Laboratory of Solid State Microstructures and school of Physics, Nanjing University, Nanjing, 210093, China}
\affiliation{Collaborative Innovation Center of Advanced Microstructures, Nanjing University, Nanjing 210093, China}
\begin{abstract}
The 1D hinge states are the hallmark of the 3D higher-order topological insulators (HOTI),
which may lead to interesting transport properties.
Here, we study the Aharonov-Bohm (AB) effect in the interferometer
constructed by the hinge states in the normal metal-HOTI junctions
with a transverse magnetic field. We
show that the AB oscillation of the conductance
can clearly manifest the spatial configurations of such hinge states.
The magnetic fluxes encircled by various
interfering loops are composed of two basic ones,
so that the oscillation of the conductance by varying the magnetic field
contains different frequency components universally related
to each other. Specifically, the
four dominant frequencies $\omega_{x,y}$ and $\omega_{x\pm y}$
satisfy the relations $\omega_{x\pm y}=\omega_x\pm\omega_y$,
which generally holds for different magnetic
field, sample size, bias voltage and weak disorder.
Our results provide a unique and robust signature of the hinge states
and pave the way for exploring AB effect in the 3D HOTI.
\end{abstract}

\date{\today}

\maketitle

\section{INTRODUCTION}
Over the past two decades, topological phases of matter
such as topological insulator and superconductor have
become an active research field of condensed matter physics \cite{RevModPhys.83.1057,RevModPhys.82.3045}.
These materials are characterized by the nontrivial band
topology and the resultant gapless (d-1)-dimensional edge states.
Very recently, the concepts of higher-order topological insulators (HOTI)
and superconductors are theoretically proposed,
which are featured by the $(d-2)$-dimensional edge states
\cite{benalcazar2017quantized,PhysRevB.96.245115,PhysRevLett.119.246402,PhysRevLett.119.246401,
schindler2018higher,PhysRevB.98.201114,PhysRevLett.123.216803,PhysRevB.97.205136,PhysRevResearch.2.043223,
PhysRevB.102.094503,PhysRevLett.123.177001,PhysRevB.99.041301,PhysRevB.100.205406,PhysRevLett.121.096803,schindler2018highernp}.
Specifically, for the 3D HOTI there exist 1D gapless states along the hinges
of the sample, so-called hinge states, while the surface
and the bulk states are both insulating.
Recent progresses have shown the evidences of the hinge
states in bismuth by the scanning-tunnelling spectroscopy
and Josephson interferometry \cite{schindler2018highernp},
which pave the way for exploring more intriguing
properties of such topological states in the HOTI.

The 1D nature of the hinge states indicates
that it is a good playground for exploring
various interference effects, such as
Aharonov-Bohm (AB) and Fabry-P\'{e}rot interferometers \cite{liang2001fabry,
PhysRevLett.62.2523,ji2003electronic,
ofek2010role,PhysRevLett.103.206806,nakamura2019aharonov}.
Actually, the chiral edge states of the quantum Hall phase
have become an important platform for the study
of mesoscopic physics, in which a variety of
novel phenomena have been observed \cite{ji2003electronic,henny1999fermionic,neder2007interference,weisz2014electronic}
due to its long coherence length and
high adjustability.
Compared with the chiral edge states,
the hinge states in the HOTI
open additional possibilities
for the implementation of novel effects due to their
3D configurations, which enrich
the way of interfering in real space.
Moreover, such effects cannot be realized in
any 2D systems, which in turn, can serve
as the deterministic evidence of the hinge states.

\begin{figure}
\centering
\includegraphics[width=0.9\columnwidth]{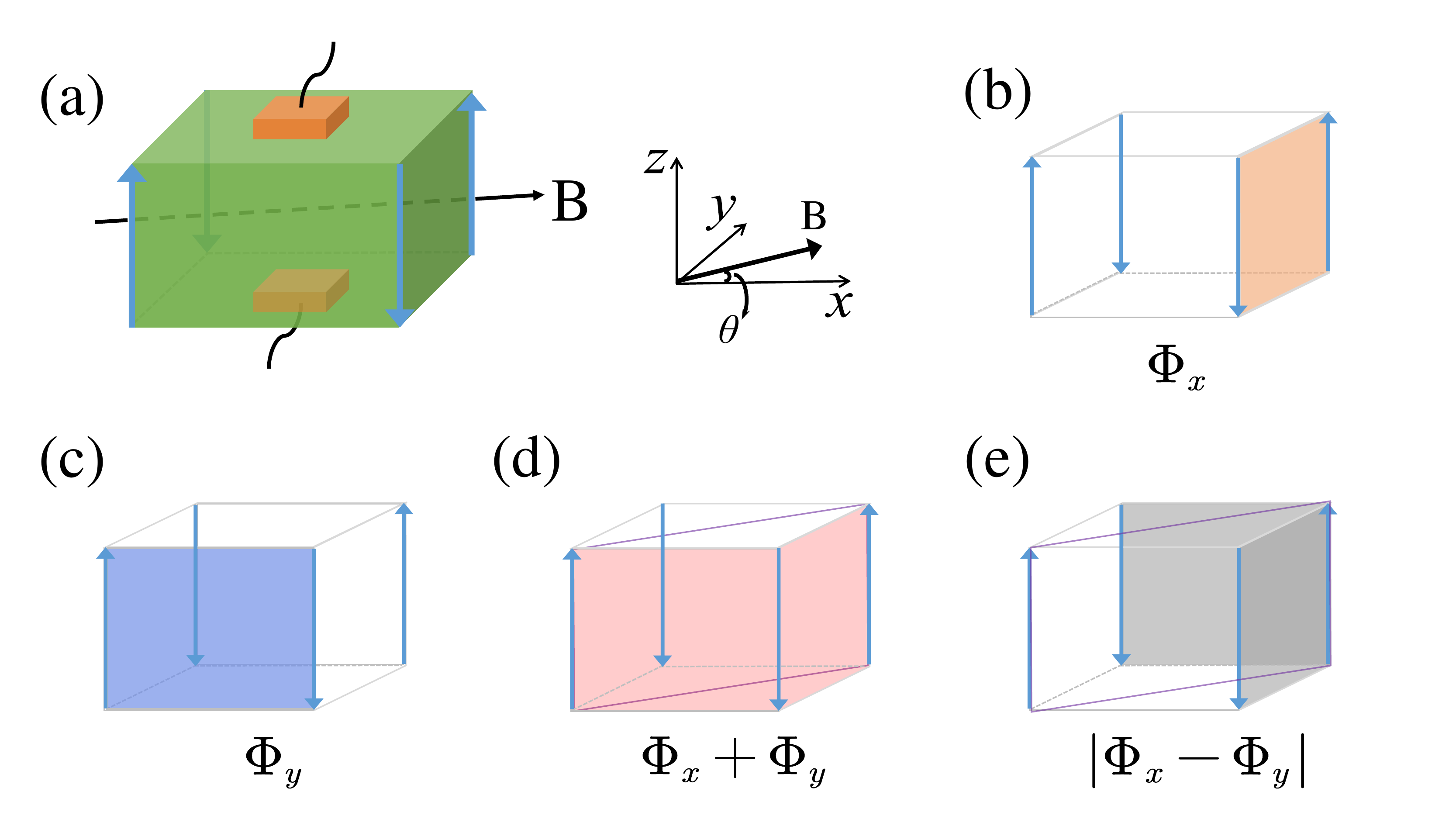}\\
\caption{(a) The interferometer
constructed by HOTI (green block) and normal metal electrodes (orange blocks).
The magnetic field $B$ is imposed in $x$-$y$ plane with a polar angle $\theta$.
(b, c) Two elemental interfering loops
with encircled flux $\Phi_x$, $\Phi_y$. (d, e) Other two dominant interfering loops
with fluxes $\Phi_x+\Phi_y$, $|\Phi_x-\Phi_y|$.
}\label{fig1}
\end{figure}

The manifestation of the AB effect in an
electron system is the periodic oscillation of conductance
as the closed trajectory of electrons encircles a magnetic flux $\Phi$ \cite{PhysRev.115.485,PhysRevLett.54.2696,PhysRevB.91.045422,PhysRevB.98.125417,PhysRevLett.114.076802,PhysRevB.74.245327}.
The dominant period of oscillation is equal
to the flux quantum $\Phi_0=h/e$, with a main
frequency $2\pi/\Phi_0$. The frequency of oscillation can be found by taking the fast Fourier
transform (FFT) of the conductance pattern \cite{PhysRevLett.54.2696,PhysRevB.91.045422,PhysRevB.98.125417}.
Recently, AB effect has been used as an effective way to detect edge states in various topological systems,
such as edge states of topological insulators \cite{peng2010aharonov,PhysRevLett.105.156803,PhysRevLett.105.206601,PhysRevB.101.045401},
Majorana fermions of topological superconductors \cite{PhysRevB.85.125440,PhysRevB.90.081405,PhysRevLett.116.166401,PhysRevResearch.2.043430}, surface states
of topological semimetal \cite{wang2016aharonov,PhysRevB.95.235436},
and non-Abelian anyons of fractional quantum Hall systems \cite{PhysRevLett.97.216404,PhysRevB.83.155440,PhysRevLett.111.186401,nakamura2019aharonov,nakamura2020direct}.

In this work, we investigate the AB effect
in the interferometer composed of the hinge
states of the quadrangular HOTI by imposing an external
magnetic field. The insulating bulk and surface states
indicate that the electron can only propagate
along the hinges of the sample, by which
the enclosed magnetic flux can lead to
a coherent oscillation of
the transmission probability.
Different from the edge states in any 2D systems,
the 3D network of the hinge states results in
peculiar interfering trajectories, which relies not only
on the magnitude of the magnetic field but also on its orientation.
The AB interferometer is sketched in Fig. \ref{fig1}(a),
where the HOTI is connected to two leads made of the normal metal
and a magnetic field $\bm{B}=(B_x,B_y)=B(\cos \theta, \sin\theta)$
is applied in the $x$-$y$ plane with $\theta$ being the polar angle.
The electrons injected from the leads propagate
along four chiral hinge states, which comprise a variety of
interfering loops; see Figs. \ref{fig1}(b)-\ref{fig1}(e).
The elemental interfering
loops shown in Figs. \ref{fig1}(b) and \ref{fig1}(c)
are exactly the boundaries of the ($\pm$1,0,0) and (0,$\pm$1,0) surfaces.
The basic loops in Figs. \ref{fig1}(b) and \ref{fig1}(c) encircle
a magnetic flux of $\Phi_x=B\cos\theta S_x$
and $\Phi_y=B\sin\theta S_y$, respectively,
with $S_{x,y}$ the surface areas.
Accordingly, the frequency components $\omega_{x}=2\pi \cos\theta S_{x}/\Phi_0$
and $\omega_{y}=2\pi \sin\theta S_{y}/\Phi_0$
naturally appear in the oscillating pattern of the conductance as
the magnetic field $B$ varies.
Interestingly, the magnetic flux in other interfering loops
can all be interpreted by the two elemental ones,
among which two typical loops in Figs. \ref{fig1}(d) and \ref{fig1}(e)
contain a flux of $\Phi_{x\pm y}=\Phi_{x}\pm\Phi_{y}$,
and the corresponding oscillating frequency components
satisfy $\omega_{x\pm y}=\omega_x\pm\omega_y$.
It turns out that the aforementioned four interfering
loops and the corresponding oscillating frequencies
dominant the coherent oscillation of the conductance.
The relation $\omega_{x\pm y}=\omega_x\pm\omega_y$
generally holds independent of various parameters such as the
magnetic field, sample size and the energy of electron, thus providing
a universal and deterministic signature of the hinge states
and HOTI.

The rest of this paper is organized as follows. In Sec. \ref{pp},
we elucidate the model of the HOTI adopted in our work.
In Sec. \ref{sm}, we apply the scattering matrix
approach to analyze the coherent transport through
the interferometer and the AB oscillation of the conductance. Detailed numerical
simulations on the lattice model are conducted in Sec. \ref{lm},
which verify the universality of the physical results.
Finally, a brief summary and outlook are given in Sec. \ref{so}.

\section{Model of HOTI}\label{pp}
We adopt the model of 3D chiral HOTI introduced by Schindler et al. \cite{schindler2018higher} as
\begin{equation}\label{hotiks}
\begin{split}
H_{\text{HOTI}}&=\left( M+t\sum_i{\cos k_i} \right) \tau_z\sigma_0 + \Delta_1\sum_i{\sin k_i \tau_x\sigma_i}\\
&\ \ \ +\Delta_2(\cos k_x - \cos k_y)\tau_y\sigma_0,
\end{split}
\end{equation}
where $\sigma_{i=x,y,z}$ and $\tau_i$ are the Pauli matrices acting on
the spin and orbital space, respectively. For $1<|M/t|<3$ and $\Delta_1,\Delta_2 \neq 0$,
the system lies in a chiral 3D HOTI phase.
The energy spectra are gapped in both the bulk and
four surfaces parallel to the $z$-axis.
Importantly, the mass term is opposite in sign
between adjacent surfaces that results in the Jackiw-Rebbi-type
bound states \cite{PhysRevD.13.3398} propagating only along the $\pm z$-direction,
or the so-called topological hinge states. Time-reversal
symmetry is broken in Eq. \eqref{hotiks}, so that the hinge states are
unidirectional or chiral, without any backscattering states
within a given hinge. Notably,
gapless Dirac cones protected by
the $\hat{C}_4^z\hat{T}$ symmetry persist on the surfaces
perpendicular to the $z$-axis \cite{schindler2018higher}.
Therefore, it is beneficial to
explore pure signature of the hinge states
through the transport in the $z$-direction.

\section{Scattering matrix analysis}\label{sm}
In this section, we study the coherent transport of electrons
through the interferometer sketched in Fig. \ref{fig1}(a)
based on
the low-energy effective model of the hinge states
using the scattering matrix approach.
The scattering matrix of the whole interferometer
can be obtained by combining those
at two normal metal-HOTI interfaces
and the matrix of phase accumulation
during propagation in the hinge states.
The matrix at the lower interface [cf. Fig. \ref{fig1}(a)] can be parameterized as
\begin{equation}\label{Sd}
S_l=\left(
    \begin{array}{cccc}
      r_1 & r_3 & t_1^{\prime} & t_3^{\prime} \\
      r_2 & r_4 & t_2^{\prime} & t_4^{\prime} \\
      t_1 & t_3 & r_1^{\prime} & r_3^{\prime} \\
      t_2 & t_4 & r_2^{\prime} & r_4^{\prime} \\
    \end{array}
  \right),
\end{equation}
which relates the incoming ($a_l$) and outgoing ($b_l$) waves
in the normal metal and the HOTI via $b_l=S_la_l$.
The matrix is assumed to be $4\times4$ such that
two incoming/outgoing waves are taken into account on both sides.
For the HOTI, the number of channels corresponds to
that of pairs of the hinge states.
The unitary condition $S_lS_l^{\dagger}=\text{I}$
is ensured by the law of current conservation.
Here,  $t_{1,\cdots,4}$ are the transmission amplitudes from the
normal metal to the chiral hinge states of the HOTI
and $r_{1,\cdots,4}$ are the corresponding reflection amplitudes.
The scattering amplitudes corresponding to
the incident waves from the hinge states of HOTI are defined
by $t^{\prime}_{1,\cdots,4}, r^{\prime}_{1,\cdots,4}$ in a similar way.
The scattering matrix for the upper interface can be defined as
\begin{equation}\label{Su}
S_u=\left(
    \begin{array}{cccc}
      r_1^u & r_3^u & t_1^{u\prime} & t_3^{u\prime} \\
      r_2^u & r_4^u & t_2^{u\prime} & t_4^{u\prime} \\
      t_1^u & t_3^u & r_1^{u\prime} & r_3^{u\prime} \\
      t_2^u & t_4^u & r_2^{u\prime} & r_4^{u\prime} \\
    \end{array}
  \right).
\end{equation}
The phase modulation of the wave function
due to the magnetic field can be described by the matrix as
\begin{equation}
S_m=\left(
  \begin{array}{cccc}
    0 & 0 & e^{i\tilde{\phi_1}} & 0 \\
    0 & 0 & 0 & e^{-i\tilde{\phi_2}} \\
    e^{-i\phi_2} & 0 & 0 & 0 \\
    0 & e^{i\phi_1} & 0 & 0 \\
  \end{array}
\right)
\end{equation}
where the phases $\phi_1, \phi_2, \tilde{\phi_1}, \tilde{\phi_2}$
are related by the magnetic fluxes through
$\phi_1+\tilde{\phi_1}=\phi_2+\tilde{\phi_2}=\phi_x =2\pi \Phi_x/\Phi_0, \phi_1-\tilde{\phi_2}=\phi_2-\tilde{\phi_1}=\phi_y=2\pi\Phi_y/\Phi_0,
\phi_1+\phi_2=\phi_{x+y}=2\pi\Phi_{x+y}/\Phi_0$
and $\tilde{\phi_1}+\tilde{\phi_2}=\phi_{x-y}=2\pi\Phi_{x-y}/\Phi_0$
with $\phi_{x,y}$ and $\phi_{x\pm y}$ being gauge invariant.

By combining three matrices $S_l, S_m, S_u$ in a standard way
we obtain the total scattering matrix for the whole system.
Here, we focus on the periods of the AB oscillation
and an overall phase shift of the pattern
is unimportant. Therefore, we can choose $S_l, S_u$ to be real for simplicity
which will not change the main results.
For an electron incident from the lower terminal,
its transmission probability $T$ to upper terminal
is obtained after some algebra as
\begin{widetext}
\begin{equation}\label{T}
\begin{split}
T&=F^{-1}\Big[C+C_X\cos\phi_x+C_Y\cos\phi_y+C_{XY}\cos\phi_{x+y}
+C_{XY}^{\prime}\cos\phi_{x-y}\Big],\\
F&=M_C+M_{XY}\cos\phi_{x+y}+M_{XY}^{\prime}\cos\phi_{x-y}+M_{2X}\cos(2\phi_x)+M_{2Y}\cos(2\phi_y)-M_X\cos\phi_x-M_Y\cos\phi_y,
\end{split}
\end{equation}
\end{widetext}
where the explicit forms of the relevant parameters are given in Appendix \ref{appA}.
The numerator of the transmission in Eq. \eqref{T}
shows that there are four dominant periodic terms
contributed by four interference loops in Figs. \ref{fig1}(b)-\ref{fig1}(e)
which correspond to four frequencies related by
$\omega_{x,y}, \omega_{x\pm y}=\omega_x\pm\omega_y$.
Note that such relations are stabilized by
the spatial configurations of the hinge states,
thus offer a clear and robust signature for its detection,
which relies little on the sample details and the
energy.
Although magnetic field in different directions
will change the values of frequencies
it does not affect the general relations between them.

\begin{figure}
\centering
\includegraphics[width=\columnwidth]{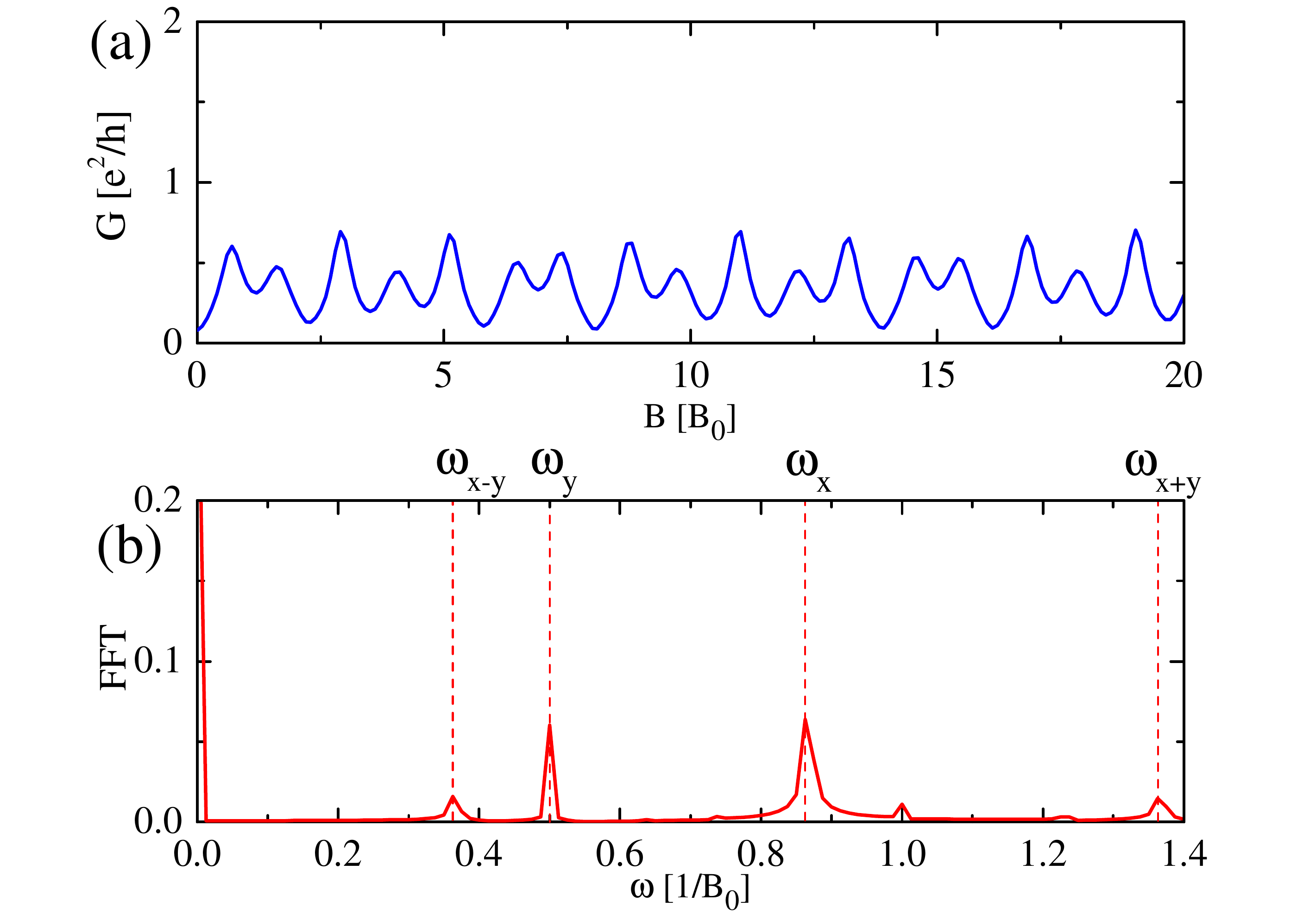}\\
\caption{(a) The conductance pattern calculated by
the scattering matrices.
(b) The FFT spectrum of the oscillation pattern.
The relevant scattering coefficients are set
to $t_1=t_3=t_4=r_1^{\prime}=r_1^u=r_2^u=\sqrt{0.4},
  t_2=r_3^{\prime}=\sqrt{0.3},
  r_1=r_2^{\prime}=\sqrt{0.2},
  r_2=r_3=r_4=t_1^u=t_2^u=t_3^u=\sqrt{0.1},
  r_4^{\prime}=t_4^u=-\sqrt{0.1},
  r_3^u=r_4^u=-\sqrt{0.4}.$}\label{fig2}
\end{figure}

Next, we provide numerical verification of
such an observation using specific scattering
amplitudes.
The AB oscillation of the conductance as a function
of $B$ and its FFT spectrum
are shown in Figs. \ref{fig2}(a) and \ref{fig2}(b), respectively.
The polar angle of the magnetic field is set to $\theta=\pi/6$ and the
unit of frequency is chosen as $1/B_0$ with $B_0=\Phi_0/(2\pi S)$
and $S=S_x=S_y$ the surface area.
One can find multiple periods in the conductance spectrum in Fig. \ref{fig2}(a).
The FFT spectrum in Fig. \ref{fig2}(b) shows that
there are four dominant frequencies with $\omega_x=0.86/B_0,
\omega_y=0.5/B_0,\omega_{x-y}=0.36/B_0,\omega_{x+y}=1.36/B_0$,
which conforms the aforementioned relation $\omega_{x\pm y}=\omega_x\pm \omega_y$.
Higher frequencies such as $\omega_{2x},\omega_{2y}$ should also appear
as that in the conventional 2D AB effect. In stark contrast,
the frequencies $\omega_{x \pm y}$ can only exist in the 3D HOTI,
which thus provides a unique evidence of the hinge states.

\section{Lattice model simulation}\label{lm}
Based on the scattering matrix analysis, we see
that there are four dominant frequencies satisfying
universal relations $\omega_{x\pm y}=\omega_x \pm \omega_y$.
In this section, we perform numerical simulation to
give rigorous results.
We write the model in Eq. \eqref{hotiks} on a cubic lattice
as \cite{PhysRevResearch.2.033327}
\begin{equation}\label{magSOTI}
\begin{aligned}
  H^{\text{Lattice}}_{\text{HOTI}}&=\sum_i{c_i^{\dagger} M\sigma_0\tau_z c_i}\\
  &+ \Bigg \{ \sum_i{{c_{i+x}^{\dagger} \bigg [ \frac{e^{i\varphi_x}}{2}(\Delta_2\sigma_0\tau_y + t\sigma_0\tau_z + i\Delta_1\sigma_x\tau_x) \bigg ] c_i}}\\
  &- \sum_i{c_{i+y}^{\dagger} (\Delta_2\sigma_0\tau_y + t\sigma_0\tau_z + i\Delta_1\sigma_y\tau_x) c_i}\\
  &+ \sum_i{c_{i+z}^{\dagger}\frac{e^{i\varphi_z}}{2}(t\sigma_0\tau_z+i\Delta_1\sigma_z\tau_x)c_i} + h.c. \Bigg \},
\end{aligned}
\end{equation}
where $c_i=(c_{a,\uparrow,i},c_{b,\uparrow,i},c_{a,\downarrow,i},c_{b,\downarrow,i})$
are the annihilate operators at lattice site $i$ with two spin
($\uparrow,\downarrow$) and two orbit ($a,b$) components.
The Peierls phase $\varphi_{x,z}=\frac{e}{\hbar} \int_{r_i}^{r_j} \bm{A}(\bm r) \cdot d\bm r$,
where $\bm{A}(\bm r)=(B_y z,0,B_x y)$ is the vector potential
under Landau gauge.
The lattice model of the normal metal electrodes is
\begin{equation}\label{NM}
\begin{aligned}
  H_{\text{NM}}&=\sum_i{(-6t+U)c_i^{\dagger}c_i}\\
  &+ \sum_i{t(c_{i+x}^{\dagger}c_i + c_{i+y}^{\dagger}c_i + c_{i+z}^{\dagger}c_i)} + h.c. .
\end{aligned}
\end{equation}

\begin{figure}
\centering
\includegraphics[width=\columnwidth]{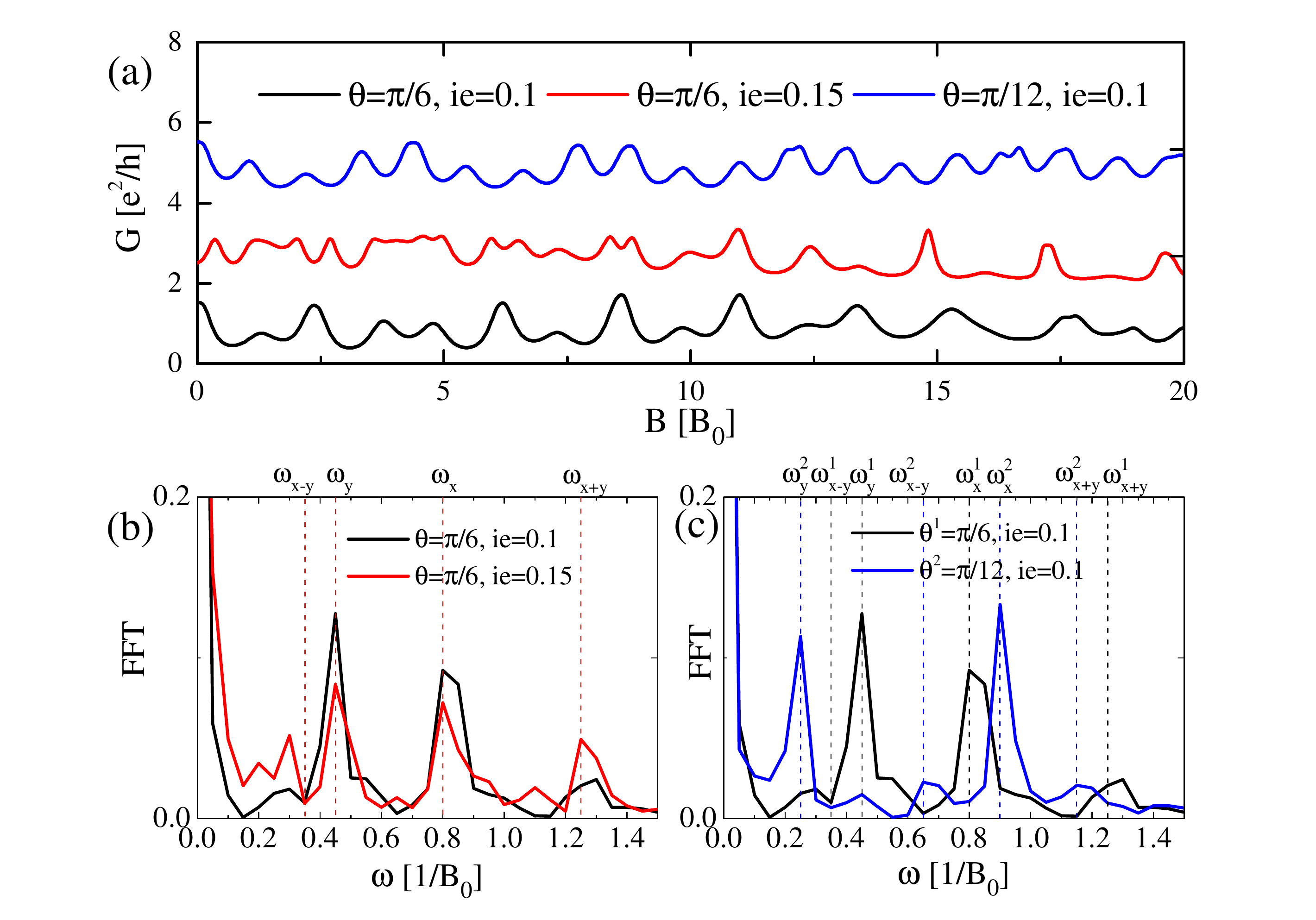}\\
\caption{(a) Conductance oscillation for different incident energy and angle $\theta$
by the lattice simulation. An offset $2e^2/h$
is imposed to adjacent curves for clarity.
(b, c) Corresponding FFT spectra of the conductance patterns.
Four dominant frequencies are marked by the dashed lines.
The model parameters are set to $t=-1, M=2.3, \Delta_1=0.8, \Delta_2=0.5, U=2$.
The lattice of the HOTI is set to a $30a\times 30a\times 30a$ cube,
where $a=1$ is the lattice constant.
}\label{fig3}
\end{figure}

The AB interferometer constructed by the 3D HOTI (green block)
and the normal metal leads (two orange blocks)
is shown in Fig. \ref{fig1}(a).
The blue arrowed lines denote the chiral hinge states.
The cross section of the HOTI
in the $x$-$y$ plane is set as $30a\times 30a$
and that for the normal metal leads is $5a\times 5a$
with $a$ being the lattice constant.
The magnetic field $B$ exists only in the HOTI region
and is parallel to the $x$-$y$ plane.

Consider an electron impinging from the normal
metal towards the HOTI with its energy lying in the
bulk gap ($\simeq$0.7$|t|$) and surface gap ($\simeq$0.32$|t|$) of the HOTI,
so that only the hinge channels are available for propagation.
Backscattering can occur at the interfaces,
giving rise to various interference loops.
The two terminal conductance $G$ is calculated
using KWANT package \cite{groth2014kwant}
and the AB conductance oscillation for different incident
energy ($ie$) and polar angle $\theta$ of magnetic field are shown in
Fig. \ref{fig3}(a) (curves are offset by $2e^2/h$ for clarity).
To get the dominant frequencies, we perform FFT
calculation whose spectra are shown in Figs. \ref{fig3}(b) and \ref{fig3}(c).

The numerical results are consistent with those by the scattering matrix
analysis in Figs. \ref{fig2}(b).
For different incident energies,
the oscillation patterns in Fig. \ref{fig3}(a) look in stark difference.
However, the dominant frequencies remain almost the same; see Fig. \ref{fig3}(b).
To check the general relations between oscillation frequencies,
we first locate two notable peaks $\omega_x$ and $\omega_y$
by the dashed lines and the other two peaks $\omega_{x\pm y}=\omega_x\pm\omega_y$
are marked accordingly in Fig. \ref{fig3}(b).
One can see that the dominant peaks match
the dashed lines very well apart from a
small deviation from $\omega_{x+y}$ for $ie=0.1$
which is attributed to the limit of numerical calculations.
For different polar angle $\theta$,
similar results can be seen in Fig. \ref{fig3}(c).
Although the location of peaks
change for different $\theta$, the general relations between them
persist.

\begin{figure}[htbp]
\centering
{
\begin{minipage}[t]{0.9\columnwidth}
\centering
\includegraphics[width=\columnwidth]{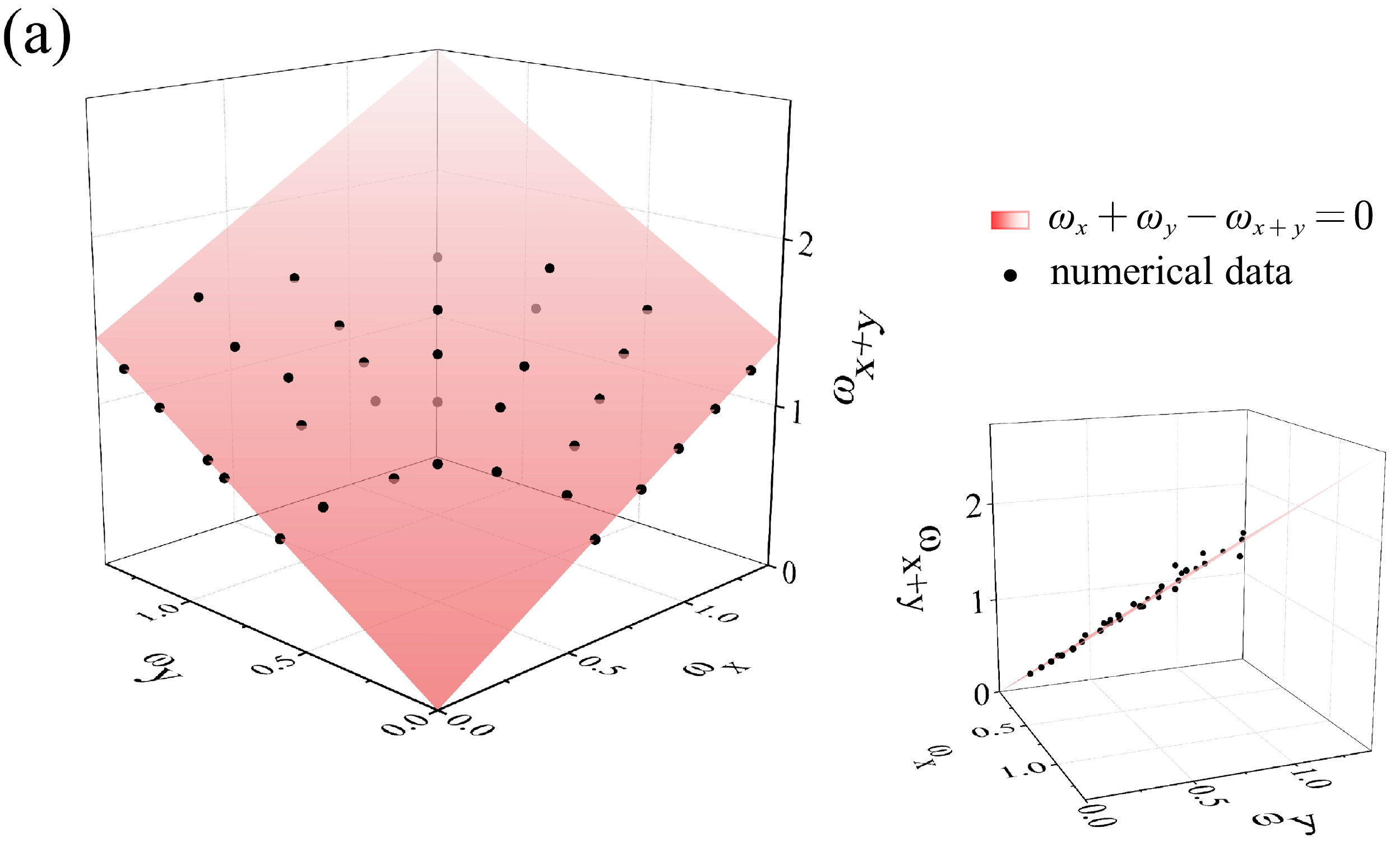}
\end{minipage}%
}
{
\begin{minipage}[t]{0.9\columnwidth}
\centering
\includegraphics[width=\columnwidth]{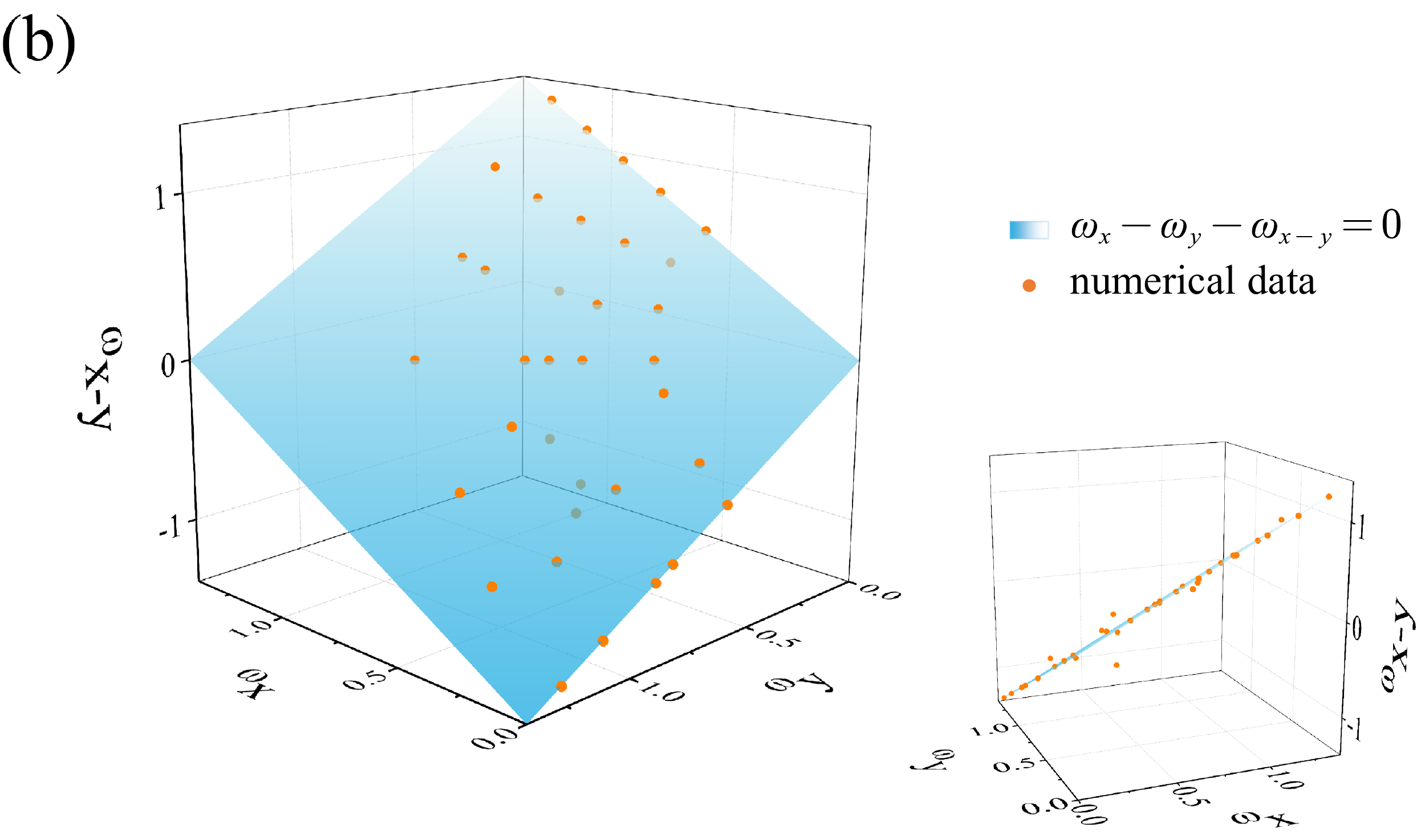}
\end{minipage}%
}
\centering
\caption{(a) $\omega_{x+y}$ and (b) $\omega_{x-y}$ as a function
of elemental frequencies $\omega_x$ and $\omega_y$. Black and
orange dots denote numerical results for different length of the HOTI
and polar angle $\theta$ of the magnetic field. The reference planes
satisfy $\omega_{x\pm y}=\omega_x\pm \omega_y$. Insets: side view
of the plots which reveal the deviation of the dots from the planes.
The parameters are the same as those in Fig. \ref{fig3}.
}
\label{fig7}
\end{figure}

In Fig. \ref{fig7}, we present more general results
by varying both the polar angle $\theta$ and the thickness
of the HOTI in the $z$-direction. Each pair of parameters
generate one point in both Figs. \ref{fig7}(a) and \ref{fig7}(b),
with its coordinates extracted in the same way as done in Fig. \ref{fig3}(b).
The reference planes therein
correspond to the frequency rule $\omega_{x\pm y}=\omega_x\pm \omega_y$.
One can see that the numerical results labeled by the
black and orange dots are well located around the reference planes,
which indicates the universality of the frequency rule.
Note that there are a few dots of negative
frequencies in Fig. \ref{fig7}(b) for $\omega_x<\omega_y$.
In experiments,
one should rather measure $|\omega_x-\omega_y|$ instead.

\begin{figure}
\centering
\includegraphics[width=\columnwidth]{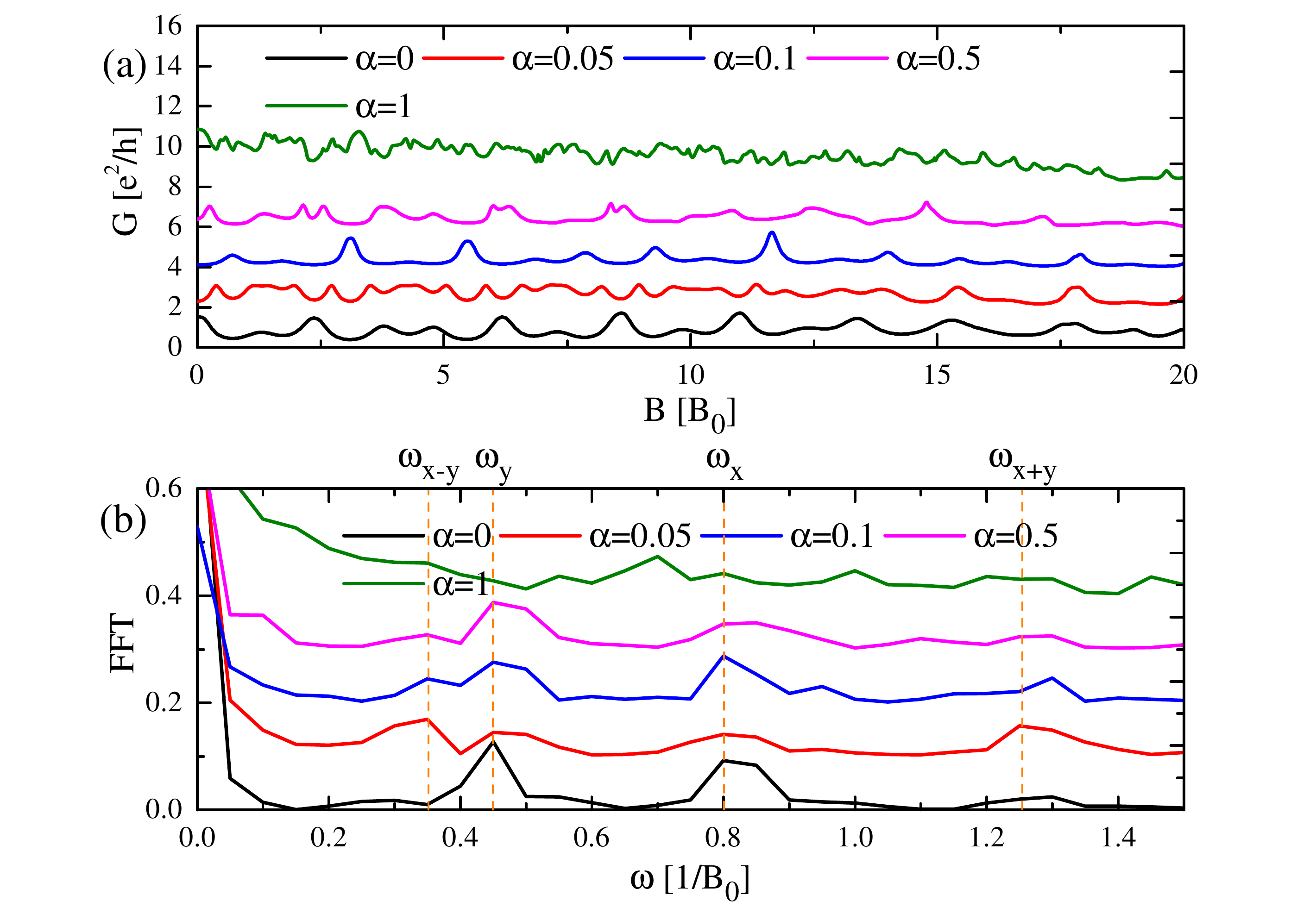}\\
\caption{(a) Conductance patterns with different
disorder strength $\alpha$.
(b) FFT spectra of the conductance.
The incident energy is $ie=0.1$ and
the polar angle of the field is $\theta=\pi/6$.
Other parameters are the same as those in Fig. \ref{fig3}.
}\label{figdis}
\end{figure}

Disorder generally exists in real samples
and it can be expected that topological chiral hinge
states and thus the AB effect should be robust,
the same as that in the quantum Hall edge states.
We show numerical results for the disorder distributed in the whole HOTI region
in Fig. \ref{figdis} with different strength $\alpha$.
For weak disorder strength $\alpha<0.7|t|$ (the gap of the bulk states),
the oscillation pattern and frequency rules $\omega_{x \pm y}=\omega_x \pm \omega_y$
retain.
For strong disorder $\alpha>0.7|t|$, the oscillation pattern
quenches stemming from disorder induced coupling
between the surface/bulk states
and the hinge states. Therefore,
as long as the disorder in the sample of HOTI
is not too strong, the AB effect can be hopefully observed.
Similar conclusion also holds for the
surface roughness.
Although the AB effect is quite robust against
the disorder effect, the observation
should be carried out within the phase coherence length
of the system. The dephasing effect always
reduces the visibility of the coherent oscillation
until it vanishes \cite{PhysRevB.75.081301,PhysRevB.98.125417}.
One more remark is that
the interference here is all of the AB type
without Al'tshuler-Aronov-Spivak (AAS) type contribution \cite{al1981aaronov,sharvin1981magnetic,al1982observation}.
The model in Eq. \eqref{hotiks} breaks time-reversal symmetry
so that the AAS effect is absent.

\section{Summary and Outlook}\label{so}

In summary, we have investigated the AB effect
in the chiral hinge states of the 3D HOTI.
Due to the spatial configurations of the hinge states,
new types of interfering loops appear
compared with the 2D AB interference.
Importantly, we predict a universal relationship
$\omega_{x \pm y}=\omega_x \pm \omega_y$ between
the dominant oscillating frequencies,
which offers a unique signature of the
hinge states as well as the HOTI.
Our study can be generalized straightforwardly to
AB effect in the 3D HOTI with helical hinge states.

\textit{Note added}.
Recently,
we became aware of a
related work \cite{PhysRevLett.127.026803},
which focuses on different aspects.

\begin{acknowledgments}
This work was supported by the National
Natural Science Foundation of
China under Grant No. 12074172 (W.C.), No. 11674160 and
No. 11974168 (L.S.), the startup
grant at Nanjing University (W.C.), the State
Key Program for Basic Researches of China
under Grants No. 2017YFA0303203 (D.Y.X.)
and the Excellent Programme at Nanjing University.
\end{acknowledgments}

\begin{appendix}

\section{specific forms of coefficient for analysis calculation}\label{appA}
In Eq. \eqref{T} of the main text, the coefficients are expressed as
\begin{widetext}
\begin{equation}
\begin{split}
C&=(t_1^2+t_3^2)(W_0^2+W_1^2+W_2^2+X_0^2+X_1^2+X_2^2)+(t_2^2+t_4^2)(Y_0^2+Y_1^2+Y_2^2+Z_0^2+Z_1^2+Z_2^2),\\
C_X&=2(W_0W_1+X_0X_1)(t_1^2+t_3^2)+2(Y_0Y_1+Z_0Z_1)(t_2^2+t_4^2)+2(W_0Y_2+W_2Y_0+X_0Z_2+X_2Z_0)(t_1t_2+t_3t_4),\\
C_Y&=2(W_0W_2+X_0X_2)(t_1^2+t_3^2)+2(Y_0Y_2+Z_0Z_2)(t_2^2+t_4^2)+2(W_0Y_1+W_1Y_0+X_0Z_1+X_1Z_0)(t_1t_2+t_3t_4),\\
C_{XY}&=2(W_0Y_0+W_0Y_0)(t_1t_2+t_3t_4),\\
C_{XY}^{\prime}&=2(W_1W_2+X_1X_2)(t_1^2+t_3^2)+2(Y1Y_2+Z_1Z_2)(t_2^2+t_4^2)
+2(W_1Y_1+W_2Y_2+X_1Z_1+X_2Z_2)(t_1t_2+t_3t_4),\\
M_{XY}&=2M_1M_4+2M_2M_3,\ \ \ \ \
M_{XY}^{\prime}=2M_1M_3+2M_2M_4,\ \ \ \ \
M_{2X}=2M_1M_2,\ \ \ \ \
M_{2Y}=2M_3M_4,\\
M_X&=2M_0M_1+2M_0M_2,\ \ \ \ \
M_Y=2M_0M_3+2M_0M_4,\ \ \ \ \
M_C=M_0^2+M_1^2+M_2^2+M_3^2+M_4^2,
\end{split}
\end{equation}
which contain the parameters defined by the elements of the
scattering matrices as
\begin{equation}
\begin{split}
W_0&=t_1^u, W_1=t_3^ur_2^{\prime}r_1^u-t_1^ur_2^{\prime}r_3^u, W_2=t_3^ur_4^{\prime}r_2^u-t_1^ur_4^{\prime}r_4^u,\\
X_0&=t_2^u, X_1=t_4^ur_2^{\prime}r_1^u-t_2^ur_2^{\prime}r_3^u, X_2=t_4^ur_4^{\prime}r_2^u-t_2^ur_4^{\prime}r_4^u,\\
Y_0&=t_3^u, Y_1=t_1^ur_3^{\prime}r_4^u-t_3^ur_3^{\prime}r_2^u, Y_2=t_1^ur_1^{\prime}r_3^u-t_3^ur_1^{\prime}r_1^u,\\
Z_0&=t_4^u, Z_1=t_2^ur_3^{\prime}r_4^u-t_4^ur_3^{\prime}r_2^u, Z_2=t_2^ur_1^{\prime}r_3^u-t_4^ur_1^{\prime}r_1^u,\\
M_0&=1+r_1^{\prime}r_1^ur_4^{\prime}r_4^u+r_3^{\prime}r_2^ur_2^{\prime}r_3^u-r_1^{\prime}r_3^ur_4^{\prime}r_2^u-r_3^{\prime}r_4^ur_2^{\prime}r_1^u,\\
M_1&=r_2^{\prime}r_3^u, M_2=r_3^{\prime}r_2^u, M_3=r_4^{\prime}r_4^u, M_4=r_1^{\prime}r_1^u.
\end{split}
\end{equation}
\end{widetext}

\end{appendix}

%\bibliography{soti-ab}

\end{document}